\documentclass[a4paper, oneside, twocolumn, notitlepage, 10pt]{extarticle_ecoc}
\usepackage{ecoc}
\usepackage{tikz}
\usepackage{adjustbox}
\usepackage{pgfplots}
\usepackage{pgfplotstable}
\usepackage{subcaption}
\usepackage{dblfloatfix}
\usepackage{afterpage}
\usepackage[compact]{titlesec}
\usetikzlibrary{arrows.meta}
\pgfplotsset{compat=newest}
\definecolor{C1}{RGB}{031, 119, 180} 
\definecolor{C2}{RGB}{255, 127, 014} 
\definecolor{C3}{RGB}{044, 160, 044} 
\definecolor{C4}{RGB}{215, 039, 040} 
\definecolor{C6}{RGB}{148, 103, 189} 
\definecolor{C100}{RGB}{140, 086, 075} 
\definecolor{C7}{RGB}{227, 119, 194} 
\definecolor{C8}{RGB}{127, 127, 127} 
\definecolor{C9}{RGB}{188, 189, 034} 
\definecolor{C5}{RGB}{023, 190, 207} 

\definecolor{C11}{RGB}{174, 199, 232} 
\definecolor{C12}{RGB}{255, 187, 120} 
\definecolor{C13}{RGB}{152, 223, 138} 
\definecolor{C14}{RGB}{255, 152, 150} 
\definecolor{C15}{RGB}{197, 176, 213} 
\definecolor{C16}{RGB}{196, 156, 148} 
\definecolor{C17}{RGB}{247, 182, 210} 
\definecolor{C18}{RGB}{199, 199, 199} 
\definecolor{C19}{RGB}{219, 219, 141} 
\definecolor{C20}{RGB}{158, 218, 229} 
\pgfplotscreateplotcyclelist{foo}{
        {C1,mark=*},
        {C2,mark=square*},
        {C3, mark = triangle*},
        {C4,mark=+},
        {C5, mark = diamond*},
        {C6, mark = x}
      }
\tikzset{>={Stealth[scale = 0.4]}}
\begin{document}
\selectlanguage{english}    


\title{
A Simplified Method for Optimising Geometrically Shaped Constellations of Higher Dimensionality}%


\author{
    Kadir G\" um\" u\c s\textsuperscript{(1)}, Bin Chen\textsuperscript{(2)},
    Thomas Bradley\textsuperscript{(1)}, Chigo Okonkwo\textsuperscript{(1)}
}

\maketitle                  


\begin{strip}
 \begin{author_descr}

   \textsuperscript{(1)} High-Capacity Optical Lab, Eindhoven University of Technology, The Netherlands,
   \textcolor{blue}{\uline{k.gumus@tue.nl}}

   \textsuperscript{(2)}
  School of Computer Science and Information Engineering, Hefei University of Technology, China
 \end{author_descr}
\end{strip}

\setstretch{1.1}
\renewcommand\footnotemark{}
\renewcommand\footnoterule{}


\begin{strip}
  \begin{ecoc_abstract}
We introduce a simplified method for calculating the loss function for use in geometric shaping, allowing for the optimisation of high dimensional constellations. We design constellations up to 12D with 4096 points, with gains up to 0.31 dB compared to the state-of-the-art. \textcopyright2023 The Author(s)
  \end{ecoc_abstract}
\end{strip}


\section{Introduction}
To increase the data rate of optical communication systems, constellation shaping in recent years has become an integral part of next-generation optical transceivers. Constellation shaping is generally divided into two subcategories, geometric (GS) and probabilistic shaping (PS). In GS, the positions of the constellation points are optimised in order to maximise the achievable information rate (AIR), while for PS the probabilities of the symbols are optimised.
\\\indent Constellation shaping has been extensively researched for both the additive white Gaussian noise (AWGN) channel \cite{Bocherer2019,Kadir2020,Frey2020,Eric2022} and non-linear fibre channel \cite{Fehenberger2016,Kojima2017,Jones2018,Liga2021,Goossens2022}, where constellations with gaps of 0.06dB\cite{Eric2022} from capacity have been designed. However, for space-division multiplexed (SDM) channels, research on constellation shaping is still very limited. In an SDM fibre, the propagated modes of the channels will experience unequal attenuation or amplification resulting in mode-dependent loss and experience interference with each other through crosstalk. Using constellation shaping, these effects could potentially be mitigated, improving the overall performance of the channel. In this paper we focus on GS, although in the future PS for use in SDM has to be looked into in order to have a proper comparison between the two.
\\\indent Note that the simplest SDM few-mode fibre, a 3-mode fibre, has 12 possible dimensions for data modulation. Previous works have been constrained to the optimisation of 2D and 4D constellations. Higher-dimensional constellations have been designed, such as in \cite{Millar:14,Rademacher2015,Bin2019,Rene2020,Mirani2021}, with constrained designs, relying on lattice-based construction, non-linear codes, or combining lower dimensional constellations respectively. Although some of these constellations are designed for SDM, they do not take into account channel-dependent effects during shaping. In order to design a constellation specifically catered towards SDM channels, where the different dimensions of the constellation interact with each other, it is desirable to design a high-dimensional constellation in an unconstrained manner.
\\\indent Unconstrained constellation design is limited to 2D and 4D because often the loss function is evaluated using the Gauss-Hermite (GH) approximation, as Monte-Carlo approximations are not suitable for shaping constellations \cite{Alex2015}. As a result, the complexity of the AIR calculations grows exponentially with the dimensionality of the constellation \cite{Alex2015}, making designing high-dimensional constellations infeasible. Designing such constellations for the AWGN channel is a first step towards constellation shaping for the SDM channel. 
\afterpage{
\setcounter{figure}{0}
\begin{figure}[t]
    \centering
      \begin{tikzpicture}
\begin{semilogyaxis}[
every axis/.append style={font=\small},
tick label style={font=\footnotesize},
xlabel = $N$,
ylabel = $L$ (------),
xmin = 2, xmax = 12,
ymin = 0.1, ymax = 1000000000000,
xtick = {2,4,8,12},
ytick = {1,100,10000,1000000,100000000,10000000000,1000000000000},
width=0.43\textwidth,
height =0.32\textwidth,
grid = major,
 xlabel near ticks,  
 ylabel near ticks,  
 xticklabel style={/pgf/number format/fixed},
every axis plot/.append style={thick},legend style={at={(0.03,0.78)},anchor=west, font = \scriptsize,row sep=-0.75ex,inner sep=0.2ex},
legend cell align={left},
]
\addplot[color = C1, mark = *]
         coordinates{(2,16)(4,128)(8,256)(12,512)};
\addlegendentry{RQ}
\addplot[color = C2, mark = square*]
         coordinates{(2,100)(4,10000)(8,100000000)(12,1000000000000)};
\addlegendentry{GH, $n = 10$}
\addplot[color = C3, mark = triangle*]
         coordinates{(2,4)(4,16)(8,256)(12,4096)};
\addlegendentry{GH, $n = 2$}
\addlegendimage{color = C2, mark = square, dashed,mark options=solid}
\addlegendimage{color = C3, mark = triangle, dashed, mark options=solid}
\addlegendentry{$R$, $n = 10$}
\addlegendentry{$R$, $n = 2$}
\end{semilogyaxis}

\begin{semilogyaxis}[
every axis/.append style={font=\small},
tick label style={font=\footnotesize},
ylabel = $R$ (--\ --\ --),
axis y line*=right,
axis x line=none,
xmin = 2, xmax = 12,
ymin = 0.1, ymax = 1000000000000,
xtick = {2,4,8,12},
ytick = {1,100,10000,1000000,100000000,10000000000,1000000000000},
width=0.43\textwidth,
height =0.32\textwidth,
 xlabel near ticks,  
 ylabel near ticks,  
 xticklabel style={/pgf/number format/fixed},
every axis plot/.append style={thick},legend style={at={(0.03,0.8)},anchor=west, font = \scriptsize,row sep=-0.75ex,inner sep=0.2ex},
legend cell align={left},
]
\addplot[color = C2, mark = square, dashed,mark options=solid]
         coordinates{(2,6.25)(4,78.125)(8,390625)(12,1953125000)};
\addplot[color = C3, mark = triangle, dashed, mark options=solid]
         coordinates{(2,0.25)(4,0.125)(8,1)(12,8)};
\end{semilogyaxis}
\end{tikzpicture}
    \vspace{-1em}
    \caption{Comparison of the amount of quadratures used between GH and RQ (16, 128, 256, 512 for 2D, 4D, 8D ,12D).}
    \label{fig:complexity}
\end{figure}
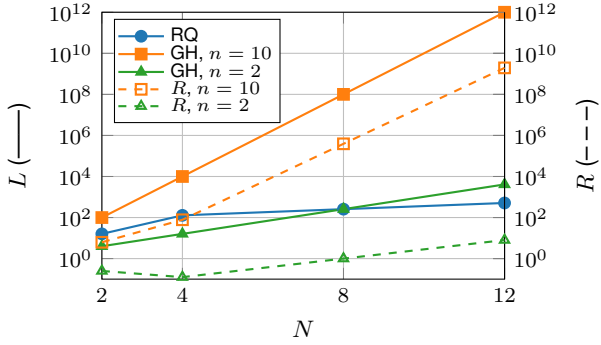}
\setcounter{figure}{1}
\begin{figure*}[b!]
    \centering
    \input{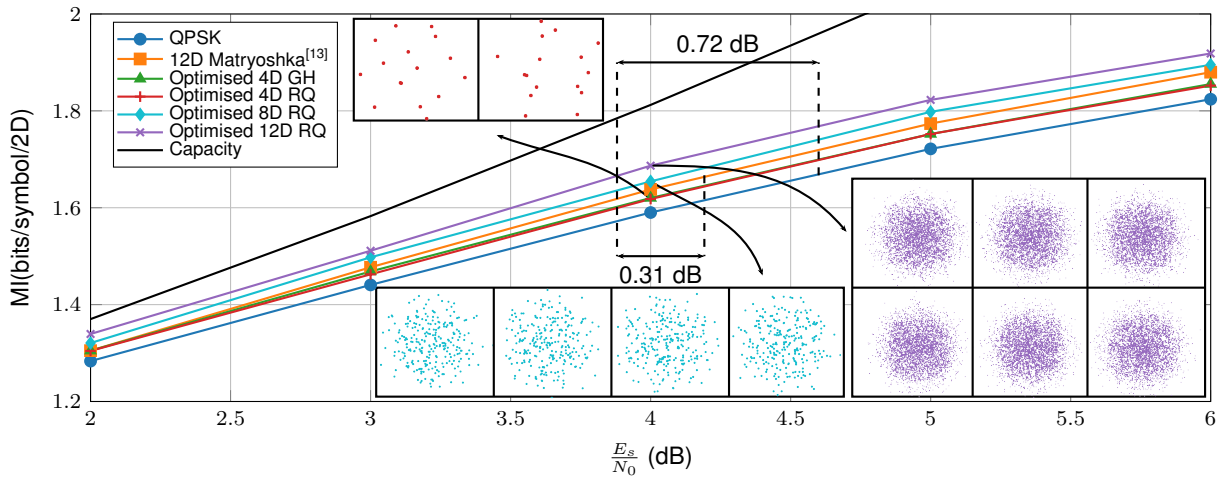}
    \vspace{-1em}
    \caption{The MI of optimised constellations of up to 12D with 2 symbols per dimension.}
    \label{fig:MI}
\end{figure*}
\\\indent In this paper, we address this first step by introducing a simplified method for geometrically shaping high dimensional constellations for both mutual information (MI) and generalised mutual information (GMI) for the AWGN channel. We optimise constellations of up to 12D with 4096 points for MI, and 12D 2048 for GMI, and show that the optimised constellations outperform state-of-the-art constellations \cite{Rene2020} by up to 0.31dB. 
\section{AIR Calculation}
The main metrics for evaluating the performance of a constellation for coded modulation systems are the MI and the GMI, where the GMI is more relevant for systems with binary forward error correction (FEC). The MI and GMI for the AWGN channel can be determined using Eqs.~(21)-(22)  from \cite{Alex2015}. 
For shaping, the MI and GMI are often evaluated using the GH approximation\cite{Alex2015}:
\begin{align}
    \int_{\mathbb{C}}e^{-||\boldsymbol{z}||^2}f(\boldsymbol{z}) &\mathrm{d}\boldsymbol{z} \approx \sum_{l_1 = 1}^n\sum_{l_2 = 1}^n\cdots\sum_{l_{N} = 1}^n\prod_{i=1}^{N}w_{GH_i}\cdot \nonumber\\
    &f(\xi_{GH_{l_1}},\xi_{GH_{l_2}},\cdots,\xi_{GH_{l_N}}),
\end{align}
where $N$ is the amount of dimensions, $n$ is the number of sample points per dimension, $\boldsymbol{w}_{GH}$ are the GH weights and $\boldsymbol{\xi}_{GH}$ are the GH quadratures. For optimisation purposes, $n$ is often chosen to be equal to 10 to give an accurate enough estimate for the optimisation\cite{Bin2023}. Using the GH approximation, the complexity grows exponentially with the dimensionality of the constellation, making optimisation of constellations beyond 4D challenging.
\\\indent We introduce a method to simplify the GH approximation to such a degree that optimising high dimensional constellations becomes feasible. Instead of using the GH quadratures, we randomly generate $L$ points in $N$ dimensional space according to the multivariate Gaussian distribution $\mathcal{N}(\boldsymbol{0},\boldsymbol{I}_N)$, and we evaluate the integral at these points. The approximation with these randomised quadratures (RQ) is:
\begin{equation}
    \int_{\mathbb{C}}e^{-||\boldsymbol{z}||^2}f(\boldsymbol{z})\mathrm{d}\boldsymbol{z} \varpropto
    \sum_{l = 1}^L w_{RQ,l} \cdot f(\boldsymbol{\xi}_{RQ,l}),
\end{equation}
where $w_{RQ,l} = e^{-\frac{||(\boldsymbol{\xi}_{RQ,l}||^2}{2}}$. It is important to note that this is not an exact approximation of the function, but instead, it provides a metric which scales with the function. This is sufficient for optimisation, as we want a metric which can act as a substitute for the AIR during the optimisation.
\\\indent In order to reduce any biases which might be introduced by an uneven sampling of the quadratures, we randomly rotate the entire set of quadratures during each optimisation iteration. Additionally, for GMI optimisations we also randomly rotate the set of quadratures for each bit. 
\\\indent By using this method, we can reduce the complexity of the loss function calculation, as shown in Fig. \ref{fig:complexity}. The number of quadratures chosen for RQ was determined empirically, as we observed that these were the minimum amount of weights required in order to have an accurate enough estimator for the optimisation. For 12D constellations, the ratio between the amount of GH and RQ quadratures $R$ is $1.95\cdot10^9$, indicating that the computation of the loss function is lower in complexity by nine orders of magnitude. Even if $n = 2$ is chosen for GH, $R$ would still be equal to 8 for 12D, while the GH approximation would not be accurate enough for use in optimisations in this case.  

\section{Results}
To verify whether the RQ approximation works for optimisation, we optimise constellations for both MI and GMI up to 12D and compare them to currently existing state-of-the-art constellations. For the optimisation, we use the same method as \cite{Kadir2020}. We focus mainly on the SNR range where the FEC overhead for a constellation is between 7 and 20\%, as this is the overhead often used in SDM systems\cite{Awaji2018}.
\setcounter{figure}{2}
\begin{figure}[!t]
    \centering
    \begin{subfigure}{2cm}
    \begin{tikzpicture}
\begin{axis}[
    width=3.6cm,
    height=4.5cm,
    xmin=-3, xmax=3,
    ymin=0, ymax=0.4,
    xticklabels = {,,},
    yticklabels = {,,},
    ticks=none
]
\pgfplotstableread{Figures/pdf_4D.txt}
\datatable
\addplot+[ybar interval,mark=no, color = C3, fill = C3, fill opacity = 0.2] table [x, y, col sep=space] {\datatable};
\pgfplotstableread{Figures/Gaussian_pdf.txt}
\datatable
\addplot[color = C4, no marks, thick]
         table
         [
          x expr=\thisrowno{0}, 
          y expr=\thisrowno{1} 
         ] {\datatable};
\end{axis}
\end{tikzpicture}
    \caption{4D}
    \end{subfigure}
    \begin{subfigure}{2cm}
    \begin{tikzpicture}
\begin{axis}[
    width=3.6cm,
    height=4.5cm,
    xmin=-3, xmax=3,
    ymin=0, ymax=0.4,
    xticklabels = {,,},
    yticklabels = {,,},
    ticks=none
]
\pgfplotstableread{Figures/pdf_8D.txt}
\datatable
\addplot+[ybar interval,mark=no, color = C5, fill = C5, fill opacity = 0.2] table [x, y, col sep=space] {\datatable};
\pgfplotstableread{Figures/Gaussian_pdf.txt}
\datatable
\addplot[color = C4, no marks, thick]
         table
         [
          x expr=\thisrowno{0}, 
          y expr=\thisrowno{1} 
         ] {\datatable};
\end{axis}
\end{tikzpicture}
    \caption{8D}
    \end{subfigure}
    \begin{subfigure}{2cm}
    \begin{tikzpicture}
\begin{axis}[
    width=3.6cm,
    height=4.5cm,
    xmin=-3, xmax=3,
    ymin=0, ymax=0.4,
    xticklabels = {,,},
    yticklabels = {,,},
    ticks=none
]
\pgfplotstableread{Figures/pdf_12D.txt}
\datatable
\addplot+[ybar interval,mark=no, color = C6, fill = C6, fill opacity = 0.2] table [x, y, col sep=space] {\datatable};
\pgfplotstableread{Figures/Gaussian_pdf.txt}
\datatable
\addplot[color = C4, no marks, thick]
         table
         [
          x expr=\thisrowno{0}, 
          y expr=\thisrowno{1} 
         ] {\datatable};
\end{axis}
\end{tikzpicture}
    \caption{12D}
    \end{subfigure}
    \vspace{0.5cm}
    \caption{The histogram of the positions of the constellation points of the 4D, 8D and 12D constellations optimised at 4dB using RQ compared to the Gaussian distribution.}
    \label{fig:pdf}
\end{figure}
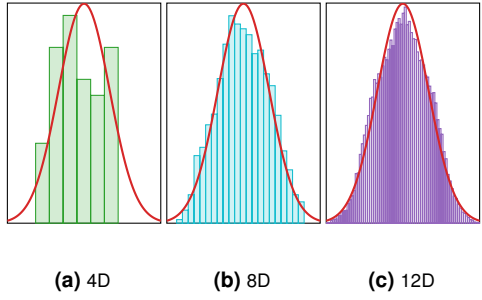
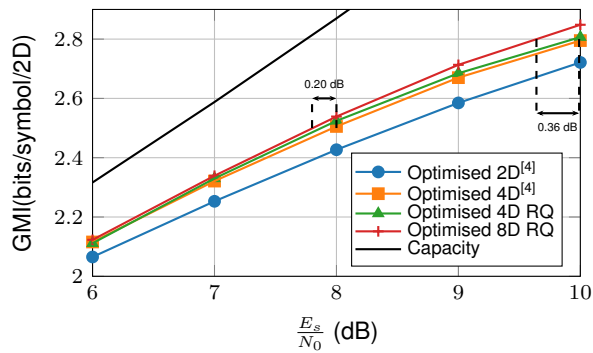
\begin{figure}[!b]
\centering  \begin{tikzpicture}
\begin{axis}[
every axis/.append style={font=\small},
tick label style={font=\footnotesize},
xlabel = $\frac{E_s}{N_0}$ (dB),
ylabel = GMI(bits/symbol/2D),
xmin = 6, xmax = 10,
ymin = 2, ymax = 2.9,
width=0.5\textwidth,
height=0.32\textwidth,
grid = major,
 xlabel near ticks,  
 ylabel near ticks,  
 xticklabel style={/pgf/number format/fixed},
every axis plot/.append style={thick},legend style={at={(0.53,0.25)},anchor=west, font = \scriptsize,row sep=-0.75ex,inner sep=0.2ex},
legend cell align={left},
cycle list name = foo,
set layers,
mark layer=axis tick labels
]

\pgfplotstableread{Figures/GMI_4_symbols_per_dimension.txt}
\datatable
\pgfplotsinvokeforeach {1,2,3,4}{
\addplot+
         table
         [
          x expr=\thisrowno{0}, 
          y expr=\thisrowno{#1} 
         ] {\datatable};
}
 \addplot[color = black, no marks]
         table
         [
          x expr=\thisrowno{0}, 
          y expr=\thisrowno{5} 
         ] {\datatable};
\addlegendentry{Optimised 2D \cite{Eric2022}}
\addlegendentry{Optimised 4D \cite{Eric2022}}
\addlegendentry{Optimised 4D RQ}
\addlegendentry{Optimised 8D RQ}
\addlegendentry{Capacity}
\draw[thick,dashed](axis cs: 9.64,2.8) -- (9.64,2.55);
\draw[thick,dashed](axis cs: 9.99,2.8) -- (9.99,2.55);
\draw[thick, <->] (axis cs:9.64,2.55) -- node[below] {\scalebox{.48} {0.36 dB}}(axis cs:9.99,2.55);
\draw[thick,dashed] (axis cs: 8,2.5) -- (8,2.6);
\draw[thick,dashed] (axis cs: 7.8,2.5) -- (7.8,2.6);
\draw[thick, <->] (axis cs:7.8,2.6) -- node[above] {\scalebox{.48} {0.20 dB}} (axis cs:8,2.6);
\end{axis}

\end{tikzpicture}
    \vspace{-1em}
    \caption{The GMI of optimised constellations of up to 8D with 8 symbols per two dimensions.}
    \label{fig:GMI_8D}
\end{figure}
\indent In Fig. \ref{fig:MI} we have optimised
4D, 8D, and 12D constellations with 2 symbols per dimension for MI. Note that randomly chosen initial constellations are used. As the figure shows, the gap to capacity shrinks as the dimensionality of the constellation increases. Figure \ref{fig:pdf} shows that as the dimensionality increases, the distribution of the constellation points more closely resembles the Gaussian distribution, hence the MI of the higher dimensional constellations is better for the AWGN channel. The 12D constellation has a gain of 0.31dB when compared to 12D Matryoshka\cite{Rene2020}, and 0.72dB gain over QPSK at the point where the FEC overhead is 20\%. 
\\\indent GMI optimisation for constellations with 8 symbols per 2 dimensions up to 8D are also performed. For the 4D constellations, the initial constellation chosen was a combination of two 2-ringed 2D APSK constellations, while for the 8D constellations, two optimised 4D constellations were combined for the initial constellation.  Fig. \ref{fig:GMI_8D} shows the GMI of these optimised constellations. Note that optimised 4D constellations outperform constellations from \cite{Eric2022}. The 8D constellations outperform the 4D constellations from \cite{Eric2022} at every SNR, with gains between 0.2 and 0.36 dB at an FEC overhead between 7\% and 20\%. 
\\\indent Finally, we have optimised a 12D constellation with 2048 points for GMI as well. We limited ourselves to 2048 points instead of 4096 because at the relevant SNR range which we are targeting (FEC overhead of 7-20\%) it is almost impossible to outperform QPSK for a constellation with the same spectral efficiency, although it is possible to create better constellations for higher SNR, as shown in \cite{Rene2020}. Therefore, optimising a constellation with 2048 points for GMI would better demonstrate the effectiveness of our proposed method. 
\\\indent The initial constellation for the optimisation is a set-partitioned (SP) 12D BPSK constellation, where the constellation consists out of all points where the $12^{th}$ bit was equal to 1, after which we removed this bit, creating a 12D constellation with 2048 points and 11 bits. Figure \ref{fig:GMI_12D} compares the GMI and normalised GMI (NGMI) of the optimised constellation to the initial constellation, QPSK and to 12D Matryoshka. It should be noted that the labeling for the 12D Matryoshka was designed for the high SNR regime, so the GMI suffers at lower SNR. The optimised constellation has a higher spectral efficiency compared to SP-12D BPSK, while being able to operate at a higher FEC rate compared to QPSK at lower SNR. 
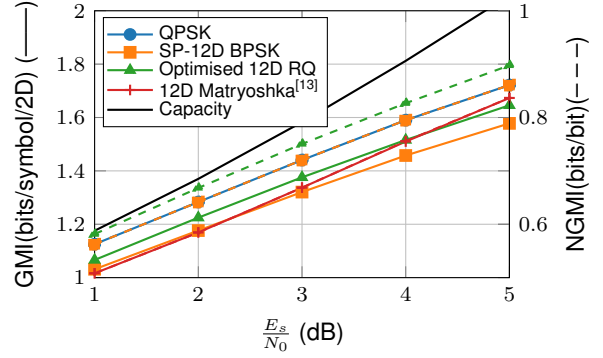
\begin{figure}[!t]
\centering  \begin{tikzpicture}
\begin{axis}[
every axis/.append style={font=\small},
tick label style={font=\footnotesize},
xlabel = $\frac{E_s}{N_0}$ (dB),
ylabel = GMI(bits/symbol/2D) (------),
xmin = 1, xmax = 5,
ymin = 1, ymax = 2,
width=0.44\textwidth,
height=0.32\textwidth,
grid = major,
 xlabel near ticks,  
 ylabel near ticks,  
 xticklabel style={/pgf/number format/fixed},
every axis plot/.append style={thick},legend style={at={(0.02,0.77)},anchor=west, font = \scriptsize,row sep=-0.75ex,inner sep=0.2ex},
legend cell align={left},
cycle list name = foo
]

\pgfplotstableread{Figures/GMI_2048_symbols_12_dimensions.txt}
\datatable
\pgfplotsinvokeforeach {1,2,3,5}{
\addplot+
         table
         [
          x expr=\thisrowno{0}, 
          y expr=\thisrowno{#1} 
         ] {\datatable};
}
 \addplot[color = black, no marks]
         table
         [
          x expr=\thisrowno{0}, 
          y expr=\thisrowno{4} 
         ] {\datatable};
\addlegendentry{QPSK}
\addlegendentry{SP-12D BPSK}
\addlegendentry{Optimised 12D RQ}
\addlegendentry{12D Matryoshka \cite{Rene2020}}
\addlegendentry{Capacity}
\end{axis}

\begin{axis}[
every axis/.append style={font=\small},
tick label style={font=\footnotesize},
xlabel = $\frac{E_s}{N_0}$ (dB),
ylabel = NGMI(bits/bit)(--\ --\ --),
xmin = 1, xmax = 5,
ymin = 0.5, ymax = 1,
axis y line*=right,
axis x line=none,
width=0.44\textwidth,
height=0.32\textwidth,
 ylabel near ticks,  
 xticklabel style={/pgf/number format/fixed},
every axis plot/.append style={thick},legend style={at={(0.5,0.2)},anchor=west, font = \scriptsize,row sep=-0.75ex,inner sep=0.2ex},
legend cell align={left},
cycle list name = foo
]

\pgfplotstableread{Figures/NGMI_2048_symbols_12_dimensions.txt}
\datatable
\pgfplotsinvokeforeach {1,2,3,4}{
\addplot+[dashed]
         table
         [
          x expr=\thisrowno{0}, 
          y expr=\thisrowno{#1} 
         ] {\datatable};
}
\end{axis}

\end{tikzpicture}
\vspace{-1em}
    \caption{The GMI and NGMI of an optimised 12D constellation with 2048 points compared to QPSK and the initial constellation SP-12D BPSK.}
    \label{fig:GMI_12D}
\end{figure}
\section{Conclusion}
In this paper, we propose a simplified method for estimating the AIR in constellation shaping, allowing for the optimisation of high dimensional constellations. We have optimised constellations for the AWGN channel for both MI and GMI up to 12D and up to 4096 and 2048 points respectively, and have shown that the optimised constellations were capable of outperforming the state-of-the-art constellations. We aim to extend this method to optimise constellations with even higher cardinality better tailored to the SDM channel.


\begin{spacing}{1}
{\footnotesize
\linespread{1} \textbf{Acknowledgements}: 
The work was the support by the Dutch Ministry of Economic Affairs and Climate Policy (EZK), as part of the Quantum Delta NL KAT-2 National Growth Funds programme on Quantum Communications and Transceiver developments in the PhotonDelta National Growth Funds programme on Integrated Photonics and  by the National Natural Science Foundation of China (No. 62171175). We gratefully acknowledge Ren\' e-Jean Essiambre for sharing data of the 12D Matryoshka constellation in \cite{Rene2020}.
}
\end{spacing}


\printbibliography

\vspace{-4mm}

\end{document}